\documentclass[preprint,showpacs,preprintnumbers,amsmath,amssymb]{revtex4}

\pdfoutput=1

\usepackage{graphicx}
\usepackage{dcolumn}
\usepackage{bm}
\usepackage{amssymb}

\begin{document}
\title{
Specific heats of quantum double-well systems 
}

\author{Hideo Hasegawa}
\altaffiliation{hideohasegawa@goo.jp}
\affiliation{Department of Physics, Tokyo Gakugei University,  
Koganei, Tokyo 184-8501, Japan}%

\date{\today}
\begin{abstract}
Specific heats of quantum systems with symmetric and
asymmetric double-well potentials have been calculated.
In numerical calculations of their specific heats,
we have adopted the {\it combined method} which takes into account
not only eigenvalues of $\epsilon_n$ for $0 \leq n \leq N_m$
obtained by the energy-matrix diagonalization but also their extrapolated ones 
for $N_m+1 \leq n < \infty$ ($N_m=20$ or 30).
Calculated specific heats are shown to be rather different 
from counterparts of a harmonic oscillator. In particular, specific heats 
of symmetric double-well systems at very low temperatures have the Schottky-type anomaly, 
which is rooted to a small energy gap in low-lying two-level eigenstates induced by a tunneling 
through the potential barrier. 
The Schottky-type anomaly is removed when an asymmetry 
is introduced into the double-well potential.
It has been pointed out that the specific-heat calculation of a double-well system
reported by Feranchuk, Ulyanenkov and Kuz'min [Chem. Phys. {\bf 157}, 61 (1991)] is misleading
because the zeroth-order operator method they adopted neglects 
crucially important off-diagonal contributions. 

\end{abstract}

        

\maketitle
\newpage
\section{Introduction}
Double-well (DW) potential models have been employed 
in a wide range of fields including physics, chemistry and biology
(for a recent review on DW systems, see Ref. \cite{Thorwart01}).
We may classify quantum DW models 
into three categories: exactly solvable, quasi-exactly solvable,
and approximately solvable models \cite{Bagchi03}. 
In the exactly solvable model, we can determine the whole spectrum 
analytically by a finite number of algebraic steps.
In contrast, we can determine a part of the whole spectrum 
in the quasi-exactly solvable model. 
In other models, eigenvalues are obtainable only by approximate, analytical or
numerical method.
Examples of exactly solvable models include the double square-well potential and 
the Manning potential \cite{Manning35}.
The Razavy potential \cite{Razavy80,Turbiner88} expressed by hyperbolic functions belongs 
to the quasi-exactly solvable models.
In this paper, we pay our attention to two types of approximately solvable models with
a quartic potential \cite{Caswell79,Balsa83,Quick85,Turbiner09} 
and a quadratic potential perturbed by a Gaussian barrier 
\cite{Chan63,Lin07}, which are hereafter referred to as model A and model B, respectively. 
These models have been commonly adopted for studies of tunneling 
and stochastic resonance in DW systems. 
It is, however, curious that studies on their thermodynamical properties are scanty 
\cite{Feynman86,Kleinert06,Okopinska87,Feranchuk91}. 
Feymann and Kleinert applied the path-integral method to a calculation 
of an effective classical partition function of DW systems \cite{Feynman86,Kleinert06}.
Okopi\'{n}ska studied the effective potential, employing the optimized and mean-field 
expansions of a path-integral representation for the partition function \cite{Okopinska87}.
The specific heat of a DW system was calculated by
Feranchuk, Ulyanenkov and Kuz'min (FUK) \cite{Feranchuk91} 
with the use of the zeroth-order operator method (ZOM) \cite{Feranchuk82}, 
related discussion being given in Sec. IV.

It is the purpose of the present paper to study the specific heat 
of quantum systems with symmetric and asymmetric DW potentials.
Various kinds of analytical and numerical methods for approximately solvable DW models 
have been proposed to evaluate their eigenvalues
\cite{Caswell79,Feynman86,Okopinska87,Feranchuk82,Balsa83,Quick85,
Turbiner09,Chan63,Feranchuk91,Lin07}.
In the present study, we evaluate them by a numerical diagonalization 
of the energy matrix with a finite size $N_m$ ($=20$ and 30).
Eigenvalues $\epsilon_n$ for $0 \leq n \leq N_m$ are sufficient for a study
of thermal properties of DW systems at very low temperature near $T=0$ K.
However, they are insufficient for describing thermodynamical properties
at elevated temperatures, as explicitly shown shortly (Figs. \ref{fig3} and \ref{fig9}).
In order to overcome this deficit, we adopt the {\it combined method} in which
we include additional eigenvalues $\epsilon'_n$, extrapolating to a larger $n$ ($N_m+1 \leq n < \infty$) 
such that they lead to results consistent with classical statistical calculations. 
Taking into account both extrapolated eigenvalues as well as those obtained 
by the energy-matrix diagonalization, we may obtain reasonable specific heats
at both low and high temperatures.

The paper is organized as follows.
In Sec. II, we will describe the adopted combined method for the energy-matrix diagonalization
and extrapolated eigenvalues. 
In Sec. III, the combined method is applied to DW systems with
a quartic potential (model A), a quadratic potential with Gaussian barrier (model B)
and the DW potential adopted by FUK (FUK model) \cite{Feranchuk91}.
In Sec. IV we critically examine a validity of the specific heat calculated by 
FUK \cite{Feranchuk91} with the use of ZOM \cite{Feranchuk82}. 
Specific heats of the triple-well system are studied also.
Sec. V is devoted to our conclusion.

\section{The combined method}
\subsection{Classical statistical calculation}
We consider a system whose Hamiltonian is given by 
\begin{eqnarray}
H &=& \frac{p^2}{2 m} + U(x),
\label{eq:A1}
\end{eqnarray}
where $m$ and $U(x)$ are a mass of a particle and a DW potential, respectively.
The classical partition function is given by
\begin{eqnarray}
Z(\beta) &=& \frac{1}{h} \int_{-\infty}^{\infty} \int_{-\infty}^{\infty} 
e^{-\beta H} \;dp\:dx
=\sqrt{\frac{m}{2 \pi \hbar^2 \beta}} \;Z_x(\beta), 
\label{eq:A5}
\end{eqnarray}
with
\begin{eqnarray}
Z_x(\beta) &=& \int_{-\infty}^{\infty} \;e^{-\beta U(x)}\;dx,
\label{eq:A5b}
\end{eqnarray}
where $\beta$ ($=1/k_B T$) denotes an inverse temperature.
From the calculated partition function $Z(\beta)$, we obtain the specific heat $C$ and 
entropy $S$ by
\begin{eqnarray}
C &=& \frac{d E}{d T}, 
\label{eq:A5c} \\
S &=& \frac{1}{T}(E-F),
\end{eqnarray}
where
\begin{eqnarray}
E &=& -\frac{\partial \ln Z(\beta)}{\partial \beta}, 
\label{eq:A5d}\\
F &=& -\frac{1}{\beta} \ln Z(\beta).
\end{eqnarray}

\subsection{Quantum statistical calculation}
In order to make a quantum statistical calculation,
it is necessary to evaluate eigenvalues of a given Hamiltonian of Eq. (\ref{eq:A1}),
for which the Shr\"{o}dinger equation is given by
\begin{eqnarray}
\left[-\frac{\hbar^2}{2m} \frac{d^2}{d x^2}+U(x) \right] \Psi(x) &=& E \Psi(x),
\end{eqnarray}
$\Psi(x)$ and $E$ standing for eigenfunction and eigenvalue, respectively.
Various approximate analytical and numerical methods have been proposed to solve
the Schr\"{o}dinger equation \cite{Caswell79,Balsa83,Quick85,Turbiner09,Chan63,Lin07}.
We evaluate eigenvalues, treating the Hamiltonian $H$ as 
\begin{eqnarray}
H &=& H_0+U(x)-U_0(x) = H_0+V(x),
\label{eq:A9}
\end{eqnarray}
with
\begin{eqnarray}
H_0 &=& 
\frac{p^2}{2m}+ U_0(x),
\label{eq:A10} \\
V(x) &=& U(x)-U_0(x), \\
U_0(x) &=& \frac{m \omega_0^2 x^2}{2},
\label{eq:A11}
\end{eqnarray}
where $\omega_0$ denotes the frequency of a harmonic oscillator.
Eigenfunction and eigenvalue for $H_0$ are given by
\begin{eqnarray}
\phi_n(x) &=& \frac{1}{\sqrt{2^n n!}} 
\left( \frac{m \omega_0}{\pi \hbar} \right)^{1/4}
\exp\left( -\frac{m \omega_0 x^2}{2 \hbar}\right)
H_n\left( \sqrt{\frac{m \omega_0}{\hbar}}\:x \right), 
\label{eq:A12}\\
E_{0n} &=& \left( n+\frac{1}{2} \right) \hbar \omega_0,
\label{eq:A13}
\end{eqnarray}
where $H_n(x)$ stands for the Hermite polynomials.
Expanding the eigenfunction $\Psi(x)$ in terms of $\phi_n(x)$
\begin{eqnarray}
\Psi(x) &=& \sum_{n=0}^{\infty} c_n \phi_n(x),
\end{eqnarray}
we obtain the secular equation for $\{ c_n \}$ expressed by
\begin{eqnarray}
E \;c_m = \sum_{n=0}^{\infty} H_{mn} c_n,
\label{eq:A14}
\end{eqnarray}
with
\begin{eqnarray}
H_{mn} &=& E_{0n}\:\delta_{mn}
+\int_{-\infty}^{\infty} \phi_m(x) V(x) \phi_n(x)\;dx,
\label{eq:A14b}
\end{eqnarray}
where $c_n$ denotes an expansion coefficient. The method mentioned above is not new,
and equivalent or similar ones have been adopted 
in Refs. \cite{Caswell79,Balsa83,Quick85,Turbiner09}.
It is possible to calculate $H_{mn}$ by MATHEMATICA \cite{MATH}.
We may diagonalize Eqs. (\ref{eq:A14}) to obtain eigenvalues $\epsilon_n$ ($n=0$ to $N_m$), 
and evaluate the quantum partition function given by
\begin{eqnarray}
Z(\beta) &=& {\rm Tr} \;e^{-\beta H}
\cong \sum_{n=0}^{N_{m}} e^{-\beta \epsilon_n},
\label{eq:A15}
\end{eqnarray}
where $N_m$ stands for the maximum eigenvalue.

As will be shown shortly (Figs. \ref{fig3} and \ref{fig9}), 
eigenvalues for $0 \leq n \leq N_m$ ($N_m \simeq 20-30$)
are sufficient for a study of low-temperature thermodynamical quantities, but
insufficient for high-temperature ones.
We adopt the combined method in which we include not only eigenvalues $\epsilon_n$ for $0 \leq n \leq N_m$
obtained by energy-matrix diagonalization but also their extrapolated ones for $n > N_m$ given by
\begin{eqnarray}
\epsilon'_n &=& A \; n^{r}\;\hbar \omega_0
\hspace{1cm}\mbox{for $N_m+1 \leq n < \infty$},
\label{eq:A16}
\end{eqnarray}
with parameters $A$ and $r$.
An exponent $r$ is chosen such that the high-temperature specific heat calculated 
with combined eigenvalues is consistent with the classical specific heat as follows.
A simple calculation of the partition function with Eq. (\ref{eq:A16}) 
in the high-temperature limit leads to
\begin{eqnarray}
Z(\beta) 
&\rightarrow& \int_{0}^{\infty} \exp[- \beta A z^r] \;dz
= (\beta A)^{-1/r} \: \Gamma\left( 1+\frac{1}{r} \right),
\label{eq:A17}
\end{eqnarray}
yielding the specific heat
\begin{eqnarray}
C &=& \left( \frac{1}{r} \right) k_B,
\label{eq:A17b}
\end{eqnarray}
which should be in agreement with the classical specific heat obtained 
by Eqs. (\ref{eq:A5c}) and (\ref{eq:A5d}).
A prefactor $A$ in Eq. (\ref{eq:A16}) is chosen such that $\epsilon'_n$ for $n \geq N_m+1$
becomes a good extrapolation of eigenvalues $\epsilon_n$ for $0 \leq n \leq N_m$
evaluated by the energy-matrix diagonalization. 
Then the resultant quantum partition function is given by
\begin{eqnarray}
Z(\beta) &=& \sum_{n=0}^{N_m} e^{-\beta \epsilon_n}
+ \sum_{n=N_m+1}^{\infty} e^{-\beta \epsilon'_n},
\label{eq:A18b}
\end{eqnarray}
where the first and second terms express contributions from
eigenvalues $\{ \epsilon_n \}$ derived by the energy-matrix diagonalization and 
from extrapolated eigenvalues $\{ \epsilon'_n \}$, respectively.

\section{Applications to model potentials}
\subsection{A quartic DW potential (model A)}
\subsubsection{The symmetric case}
\begin{figure}
\begin{center}
\includegraphics[keepaspectratio=true,width=80mm]{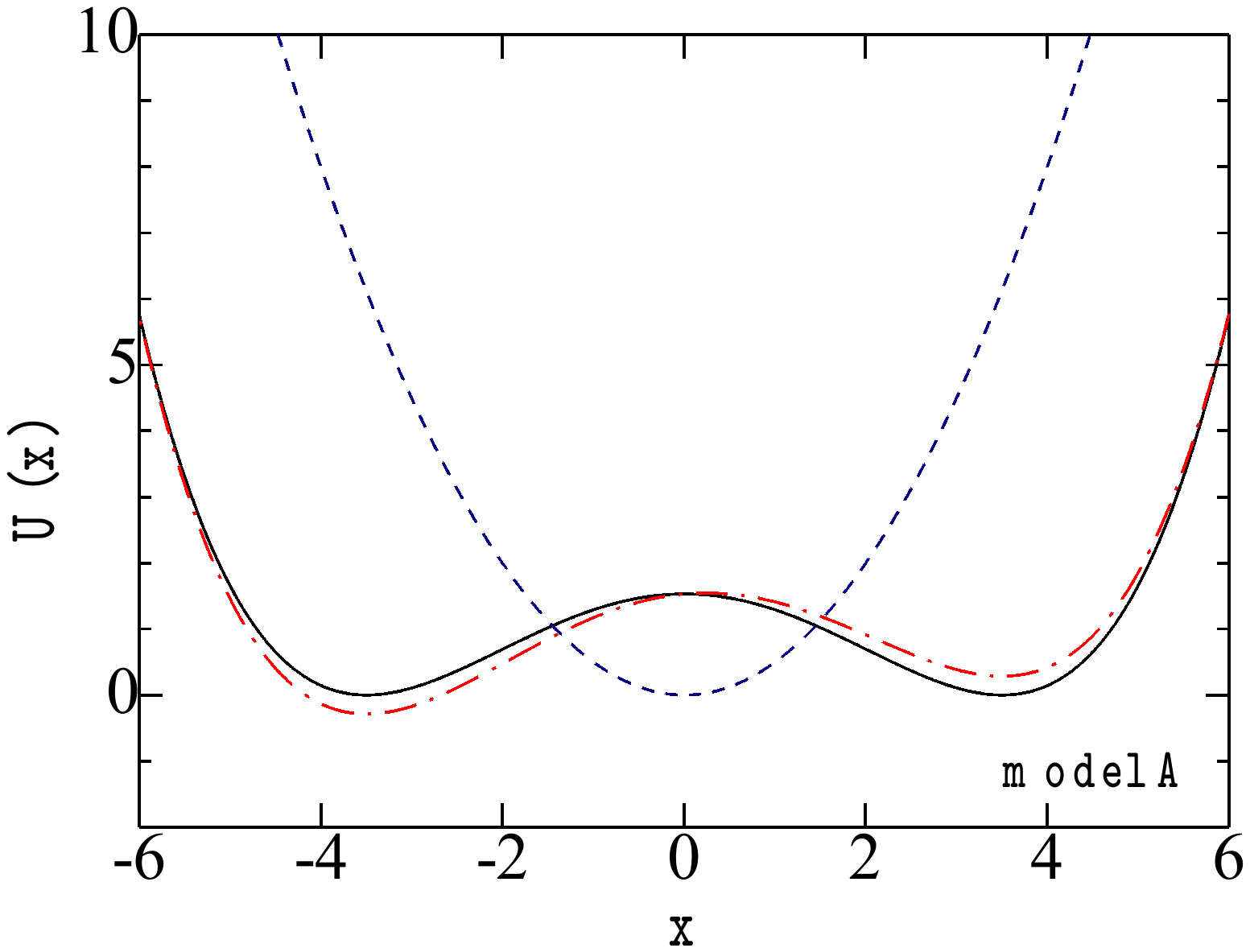}
\end{center}
\caption{
(Color online) 
The symmetric quartic potential [Eq. (\ref{eq:A4})] (solid curve) 
and asymmetric ones  [Eq. (\ref{eq:C1})] with $d=0.01$ (chain curve)
and $d=-0.01$ (double chain curve) of model A, 
the dashed curve expressing the harmonic potential [Eq. (\ref{eq:A11})].
}
\label{fig1}
\end{figure}

First we apply our combined method to model A with
the symmetric DW potential given by
\begin{eqnarray}
U(x) &=& \frac{m \omega_0^2}{8 x_0^2}\;(x^2-x_0^2)^2,
\label{eq:A4}
\end{eqnarray}
which has stable minima at $x= \pm x_0$ and an unstable maximum at $x=0$.
The height of the potential barrier is $\Delta=U(0)-U(\pm x_0)$ 
with $U(0)=m \omega_0^2 x_0^2/8$ and $U(\pm x_0)=0$.
For a later purpose of an energy-matrix calculation,
we have chosen $U''(\pm x_0)=m \omega_0^2$ such that the potential $U(x)$ has the
same curvatures at the minima as the harmonic potential $U_0(x)$ given by Eq. (\ref{eq:A11}).
For numerical calculations, we assume $m=1.0$, 
$\omega_0=1.0$ and
$x_0=3.5$ \cite{MATH}, for which $U^{''}(x_0)=U_0^{''}(0)=1.0$
and $\Delta=1.531$.
The symmetric DW potential in model A is plotted by the solid curve
in Fig. \ref{fig1}, where the harmonic potential given by Eq. (\ref{eq:A11}) is shown 
by the dashed curve: chain and double-chain curves will be explained later [Eq. (\ref{eq:C1})].

\begin{figure}
\begin{center}
\includegraphics[keepaspectratio=true,width=100mm]{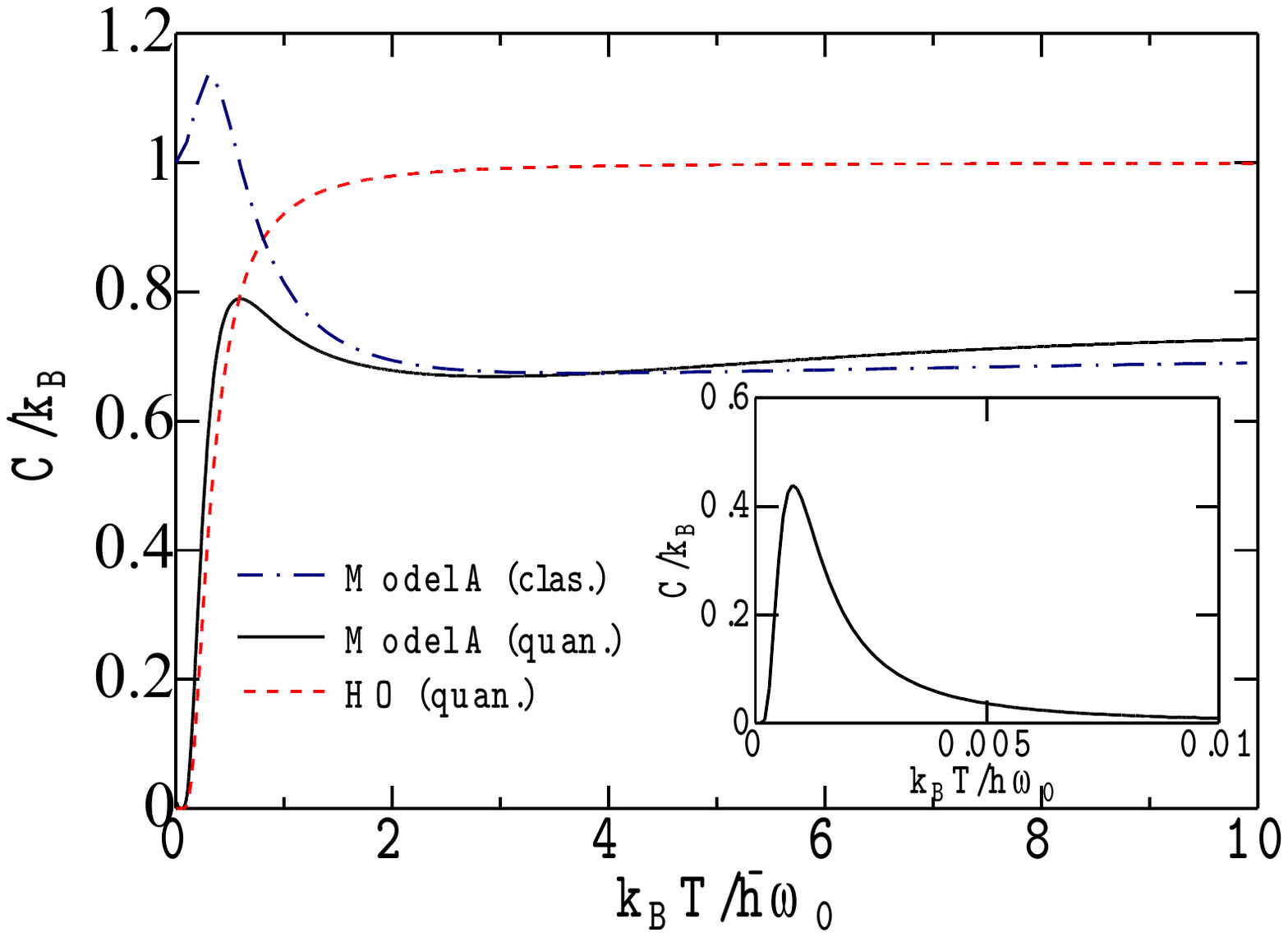}
\end{center}
\caption{
(Color online) 
Temperature dependences of
classical (chain curve) and quantum specific heats (solid curve) of model A
with the symmetric DW potential [Eq. (\ref{eq:A4})],
dashed curve expressing quantum specific heat of a harmonic oscillator (HO).
The inset shows an enlarged plot of the quantum specific heat
at very low temperatures with the Schottky-type anomaly.
}
\label{fig2}
\end{figure}

We have calculated the classical specific heat which is shown by the chain curve 
in Fig. \ref{fig2}. 
The calculated specific heat in the high-temperature limit of $T \rightarrow \infty$ reduces to
\begin{eqnarray}
C &=&\left( \frac{1}{2}+\frac{1}{4} \right) \:k_B =  \frac{3}{4} \;k_B,
\label{eq:A6}
\end{eqnarray}
because we obtain $C(T)/k_B=0.690$, 0.7293, 0.7434 and 0.7473 for 
$k_B T/\hbar \omega_0=10.0$, 100.0 and 1000.0,
respectively. 
The first (1/2) and second (1/4) terms in Eq. (\ref{eq:A6}) express contributions from
momentum ($p$) and coordinate ($x$), respectively, the latter being due to
the quartic power of the potential. Indeed, in a system with a quartic potential of 
$U(x)=x^4/4$, the coordinate contribution to the classical specific heat becomes 
$(1/4) k_B$ (the Virial theorem).

For a special case of the DW potential
\begin{eqnarray}
U(x) &=& \frac{x^4}{4}-\frac{x^2}{2}, 
\label{eq:A7}
\end{eqnarray}
we obtain the analytical expression for $Z_x(\beta)$,
\begin{eqnarray}
Z_x(\beta) &=&  \left( \frac{\pi}{2} \right)\; e^{\beta/8}
\left[ I_{-\frac{1}{4} }\left (\frac{\beta}{8} \right) 
+ I_{\frac{1}{4}}\left( \frac{\beta}{8} \right) \right],
\label{eq:A8}
\end{eqnarray}
which yields $C(T)/k_B=0.7236$, 0.7416, 0.7473 for $T=10$, 100 and 1000,
respectively, $I_n(z)$ denoting the modified Bessel function of the first kind.

\begin{figure}
\begin{center}
\includegraphics[keepaspectratio=true,width=90mm]{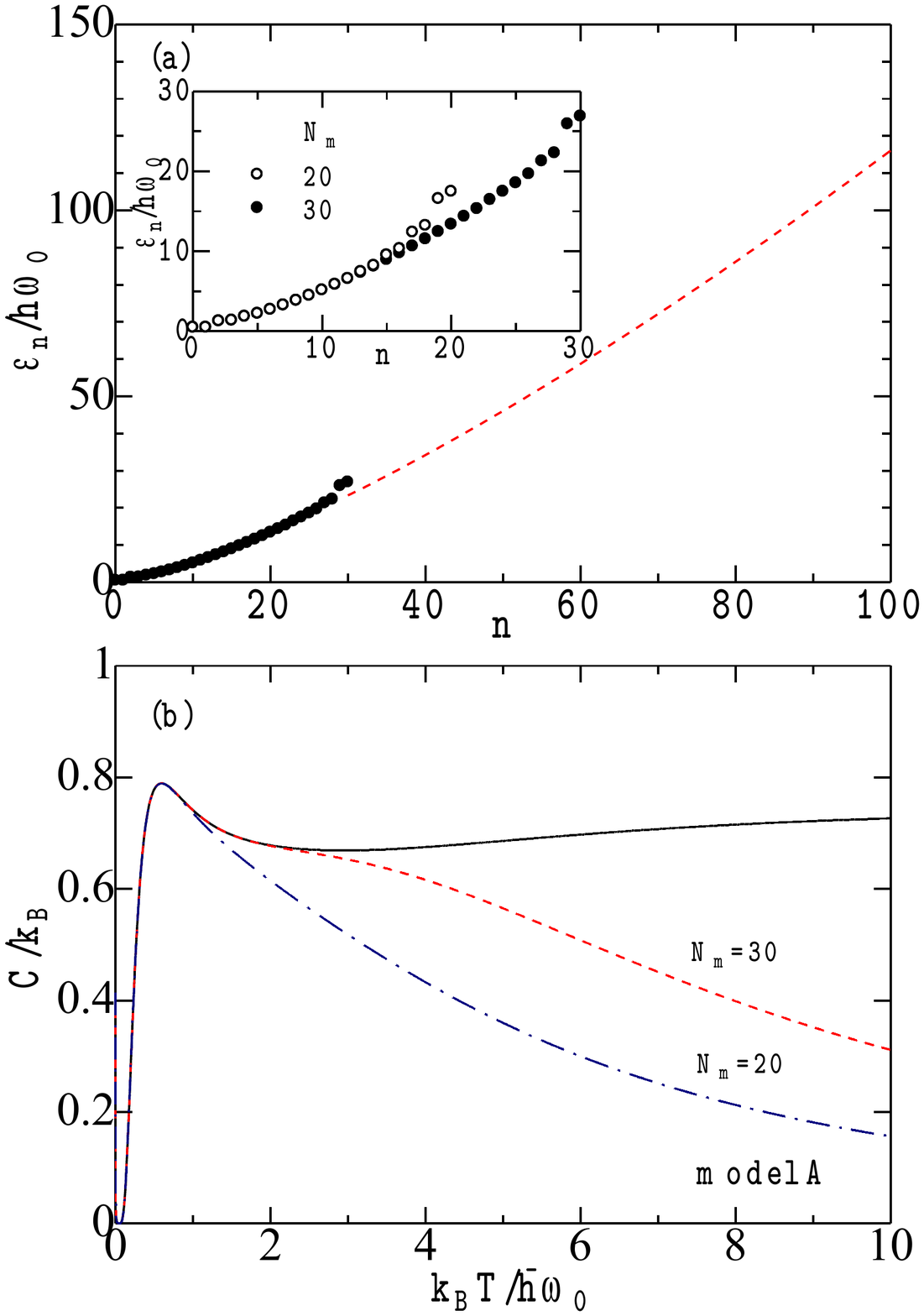}
\end{center}
\caption{
(Color online) 
(a) The $n$ dependence of eigenvalues $\epsilon_n$ of model A with the symmetric DW potential 
[Eq. (\ref{eq:A4})] obtained by the energy-matrix diagonalization 
for $0 \leq n \leq N_m$ (circles) and by an extrapolation given by
$\epsilon'_n=0.25 \: n^{4/3} \;\hbar \omega_0$ for $n > N_m$ (dashed curve),
the inset showing eigenvalues obtained 
by the energy-matrix diagonalization with $N_m=20$ (open circles) and $N_m=30$ (filled circles).
(b) The temperature dependence of the quantum specific heat calculated 
with eigenvalues $\epsilon_n$ for $0 \leq n \leq N_m$ with $N_m=20$ (chain curve)
and $N_m=30$ (dashed curve), the solid curve expressing the result with
combined eigenvalues shown by the dashed curve in (a).
}
\label{fig3}
\end{figure}

Matrix elements $H_{mn}$ of Eq. (\ref{eq:A14b}) are finite for pairs of 
$\vert m-n \vert =0$, $2$ and $4$.
Figure \ref{fig3}(a) shows eigenvalues $\{\epsilon_n \}$ obtained for $N_m=20$ (open circles) 
and 30 (filled circles).
Eigenvalues for $n < 20 $ are almost the same for $N_m=20$ and 30.
Calculated eigenvalues $\epsilon_n$ for $n=0$, 1, 2 and 3 are 
0.476188, 0.478131, 1.2695 and 1.3514, respectively, with $N_m=30$. 
Eigenvalues of $\epsilon_0=0.476188$ and $\epsilon_1=0.478131$ originate
from two eigenvalues of $E_{00}=0.5$ in Eq. (\ref{eq:A13}) at two minima 
with symmetric and antisymmetric wavefunctions.
They are quasi-degenerated states with a small gap given by 
$\delta \equiv  \epsilon_1-\epsilon_0 = 0.001943$,
which is induced by the tunneling effect through the potential barrier.
Similarly, eigenvalues for $n=2$ and 3 are also quasi-degenerated
as given by $\epsilon_3-\epsilon_2 = 0.0819$.

\begin{figure}
\begin{center}
\includegraphics[keepaspectratio=true,width=100mm]{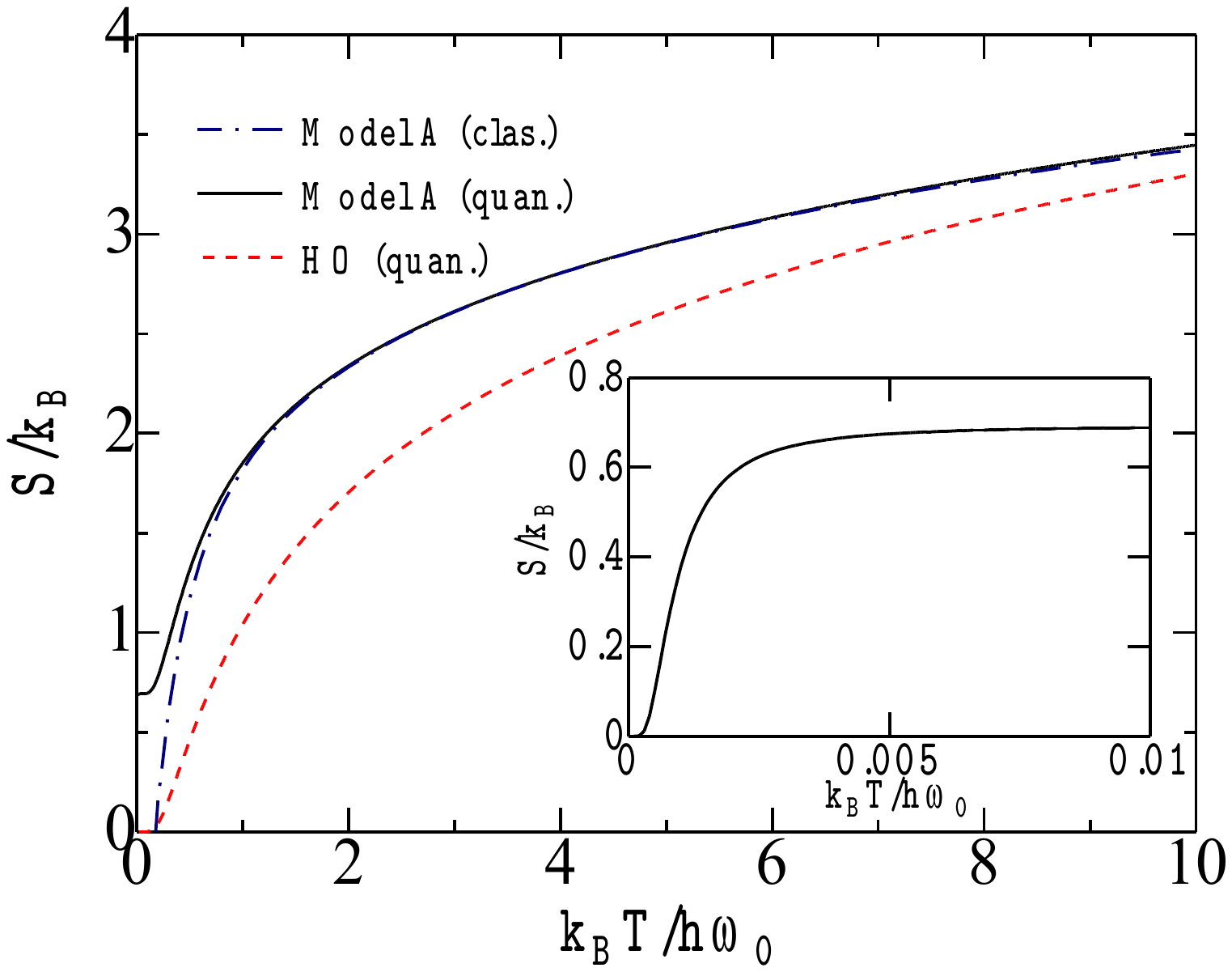}
\end{center}
\caption{
(Color online) 
Temperature dependences of classical (chain curve) and quantum entropies (solid curve) 
of model A with the symmetric potential [Eq. (\ref{eq:A4})],
and the quantum entropy of a harmonic oscillator (HO) (dashed curve),
the inset showing an enlarged plot of the quantum entropy at very low temperatures.
}
\label{fig4}
\end{figure} 
 
Specific heats calculated with the use of these eigenvalues $\epsilon_n$ for $0 \leq n \leq N_m$
with $N_m=20$ and 30 are plotted by dashed and chain curves, respectively,
in Fig. \ref{fig3}(b). They are in good agreement at $k_B T/\hbar \omega_0 < 0.5$
but significantly different at $k_B T/\hbar \omega_0 > 1.0$.
This implies that eigenvalues for $0 \leq n \leq N_m$ ($=20$ and 30) are insufficient for a study of
the specific heat at elevated temperatures of $k_B T/\hbar \omega_0 > 1.0$.

We adopt the combined method with extrapolated eigenvalues given by
\begin{eqnarray}
\epsilon'_n &=& 0.25 \; n^{4/3}\;\hbar \omega_0
\hspace{1cm}\mbox{for $N_m+1 \leq n < \infty$},
\label{eq:A16B}
\end{eqnarray}
where an exponent of $r=4/3$ chosen by $C/k_B=1/r=3/4$ in Eqs. (\ref{eq:A17b}) and (\ref{eq:A6}) is consistent with the WKB type analysis for a large $n$.
Extrapolated eigenvalues given by Eq. (\ref{eq:A16B}) are plotted in Fig. \ref{fig3}(a). 
The quantum specific heat of model A calculated with combined eigenvalues
is shown by the solid curve in Fig. \ref{fig2} [or Fig. \ref{fig3}(b)] \cite{Note3}.
The quantum specific heat is rather different from the classical one at low temperatures as expected.
A closer inspection of the quantum specific heat reveals that
$C(T)$ has an anomalous peak at very low temperature at 
$k_B T/\hbar \omega_0 \simeq 0.001 \sim \delta/2$, as shown in the inset of Fig. \ref{fig2}.
It is the Schottky-type specific heat arising from low-lying two-level eigenvalues
of $\epsilon_0$ and $\epsilon_1$ whose energy gap is induced by a mixing through a tunneling.
Although quantum and classical specific heats do not well agree 
at $k_B T/\hbar \epsilon_0 \sim10$ in Fig. \ref{fig2},
both reduce to $(3/4) k_B$ in the high-temperature limit

For a comparison, we show the quantum specific heat of a harmonic oscillator 
by the dashed curve in Fig. \ref{fig2}. 
The quantum specific heat of model A
is not dissimilar to that of a harmonic oscillator at low temperatures 
except for the Schottky-type anomaly. 
However, the high-temperature specific heat of model A
given by $C=(3/4) k_B$ is different from $C_{HO}=k_B$ of a harmonic oscillator.

Temperature dependences of classical and quantum entropies of model A are shown 
by chain and solid curves, respectively, in Fig. \ref{fig4} where
the quantum entropy of a harmonic oscillator is plotted by the dashed curve.
With decreasing the temperature, the quantum entropy
decreases but seems to remain at $0.69$ ($\simeq \ln2$). The inset of Fig. \ref{fig4} shows
that it furthermore decreases below $k_B T/\hbar \omega_0 \simeq 0.002$
and approaches zero at vanishing temperature in consistent with the third thermodynamical law.
This rapid change of the entropy is related with the Schottky-type specific heat 
shown in the inset of Fig. \ref{fig2}.

\subsubsection{The asymmetric case}
Next we apply the combined method to model A with the asymmetric DW potential given by
\begin{eqnarray}
U(x) &=& \frac{m \omega_0^2}{8 x_0^2}\;(x^2-x_0^2)^2 
- d \left( \frac{x^3}{3}- x_0^2 x \right),
\label{eq:C1}
\end{eqnarray}
where $d$ signifies a degree of the asymmetry.
Locally-stable minima of the potential 
locate at $x= \pm x_0$ and an unstable maximum is at $x_u = d \:(2 x_0^2/m \omega_0^2)$ with
\begin{eqnarray}
U(\pm x_0) &=& \pm \:\frac{2 d x_0^3}{3}, \\
U(x_u) &=& \frac{m \omega_0^2 x_0^2}{8}+ \frac{d^2 x_0^2}{ m \omega_0^2} 
- \frac{2 d^4  x_0^4}{3 m^3 \omega_0^6}, \\
\Delta U &=& U(x_0)-U(- x_0)=\frac{4 d x_0^3}{3}.
\end{eqnarray}
The asymmetry parameter $d$ is assumed to be given by
\begin{eqnarray}
-d_c < d < d_c = \frac{m \omega_0^2}{2 x_0},
\end{eqnarray}
for which $x_u$ locates at $-x_0 < x_u < x_0$.
We obtain $d_c=1/7$ for adopted parameters of $m=1.0$, $\omega_0=1.0$ and $x_0=3.5$.
In the limit of $d=0$, $U(x)$ in Eq. (\ref{eq:C1}) reduces to the symmetric DW potential
given by Eq. (\ref{eq:A4}).

Table 1 shows potential values of $U(-x_0)$, $U(x_u)$, $U(x_0)$
and $\Delta U$ as a function of $d$.
When a sign of $d$ is changed, those of $U(-x_0)$, $U(x_0)$ and $\Delta U$
are changed, but $U(x_u)$ is unchanged.
With increasing $\vert d \vert $, $\vert \Delta U \vert$ is gradually increased. 
Chain and double-chain curves in Fig. \ref{fig1} show $U(x)$
for $d=0.01$ and $-0.01$, respectively, for which a difference of $\vert \Delta U \vert$ 
is about 30 \% of the potential barrier of $\vert U(x_u)-U(-x_0) \vert$.

\begin{figure}
\begin{center}
\includegraphics[keepaspectratio=true,width=100mm]{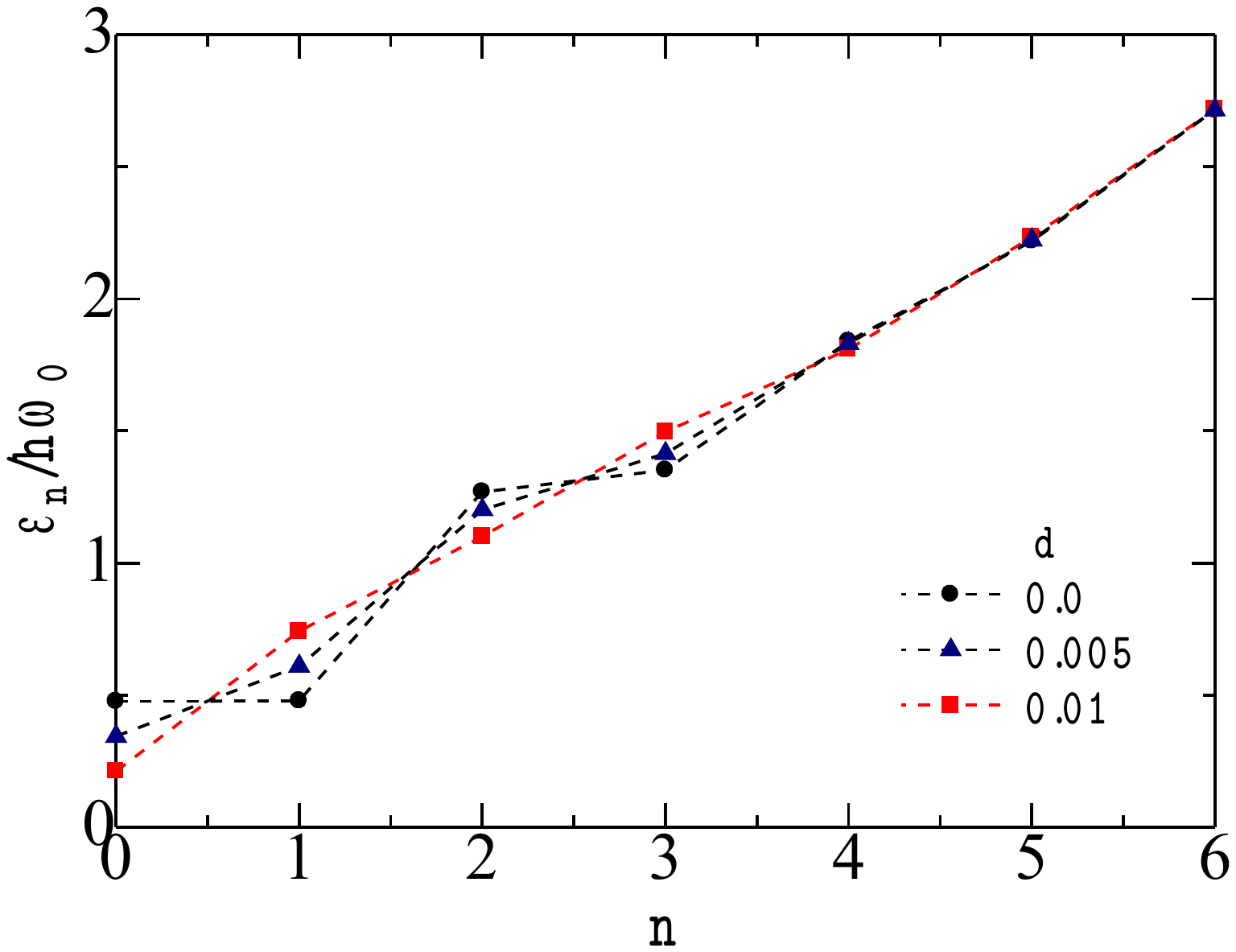}
\end{center}
\caption{
(Color online) The $n$ dependence of eigenvalues $\epsilon_n$ of model A 
with the asymmetric DW potential [Eq. (\ref{eq:C1})] 
for various $d$: $d=0.0$ (circles), 0.005 (triangles) and 0.01 (squares) 
with $N_m=30$, dashed curves being plotted for guide of the eye.
}
\label{fig5}
\end{figure}

Matrix elements $H_{mn}$ of Eq. (\ref{eq:A14b}) are not vanishing for pairs of $\vert m-n \vert \leq 4$.
Eigenvalues for $n=0$ and 1 are quasi-degenerated for $d=0.0$ as mentioned before.
This quasi-degeneracy is removed with an introduction of $d$: 
$\delta$ ($= \epsilon_1-\epsilon_0$) is increased 
with increasing $\vert d \vert$ as shown in Table 1.
Eigenvalues $\epsilon_n$ for $d=0.0$ (circles), 0.005 (triangles) and 0.01 (squares) 
evaluated by the energy-matrix diagonalization with $N_m=30$ are plotted as a function of $n$
in Fig. \ref{fig5}, where an increase in $\delta$ 
with increasing $d$ is clearly realized.

\begin{center}
\begin{tabular}[t]{|c|c|c|c|c|c|}
\hline
$d$ & $U(-x_0)$ & $U(x_u)$ & $U(x_0)$ &  $\Delta U$ & $\delta$
\\ \hline \hline
0.0 & 0.0  & $\;\;1.53125\;\;$ &  0.0 &  0.0 & $\;\;0.001943\;\;$ \\
$\;\pm 0.001\;$ & $\mp 0.02858$ & $\;1.53140\;$  &  $\pm 0.02858$ & $\pm 0.057166$ & $\;0.053123\;$  \\
$\pm 0.002$ & $\mp 0.05716$ & 1.53185 &  $\pm 0.05716$ & $\pm 0.11433$ & 0.10619  \\
$\pm 0.003$ & $\mp 0.08575$ & 1.53260 &  $\pm 0.08575$ & $\pm 0.16184$ & 0.15927  \\
$\pm 0.004$ & $\mp 0.11433$ & 1.53337 &  $\pm 0.11433$ & $\pm 0.22867$ & 0.21235  \\
$\pm 0.005$ & $\mp 0.14292$ & 1.53500 &  $\pm 0.14292$ & $\pm 0.28583$ & 0.26542    \\
$\pm 0.01$  & $\mp 0.28583$ & 1.54624 &  $\pm 0.28583$ & $\pm 0.57167$ & 0.53070  \\
\hline
\end{tabular}
\end{center}

{\it Table 1} Potential values at locally-stable minima ($\pm x_0$), 
an unstable maximum position ($x_u$), $\Delta U$ [$= U(x_0)-U(-x_0)$],
and the energy gap $\delta$ ($= \epsilon_1-\epsilon_0$) 
as a function of the asymmetry $d$ of model A with the asymmetric potential 
$U(x)$ [Eq. (\ref{eq:C1})] ($N_m=30$).

\begin{figure}
\begin{center}
\includegraphics[keepaspectratio=true,width=100mm]{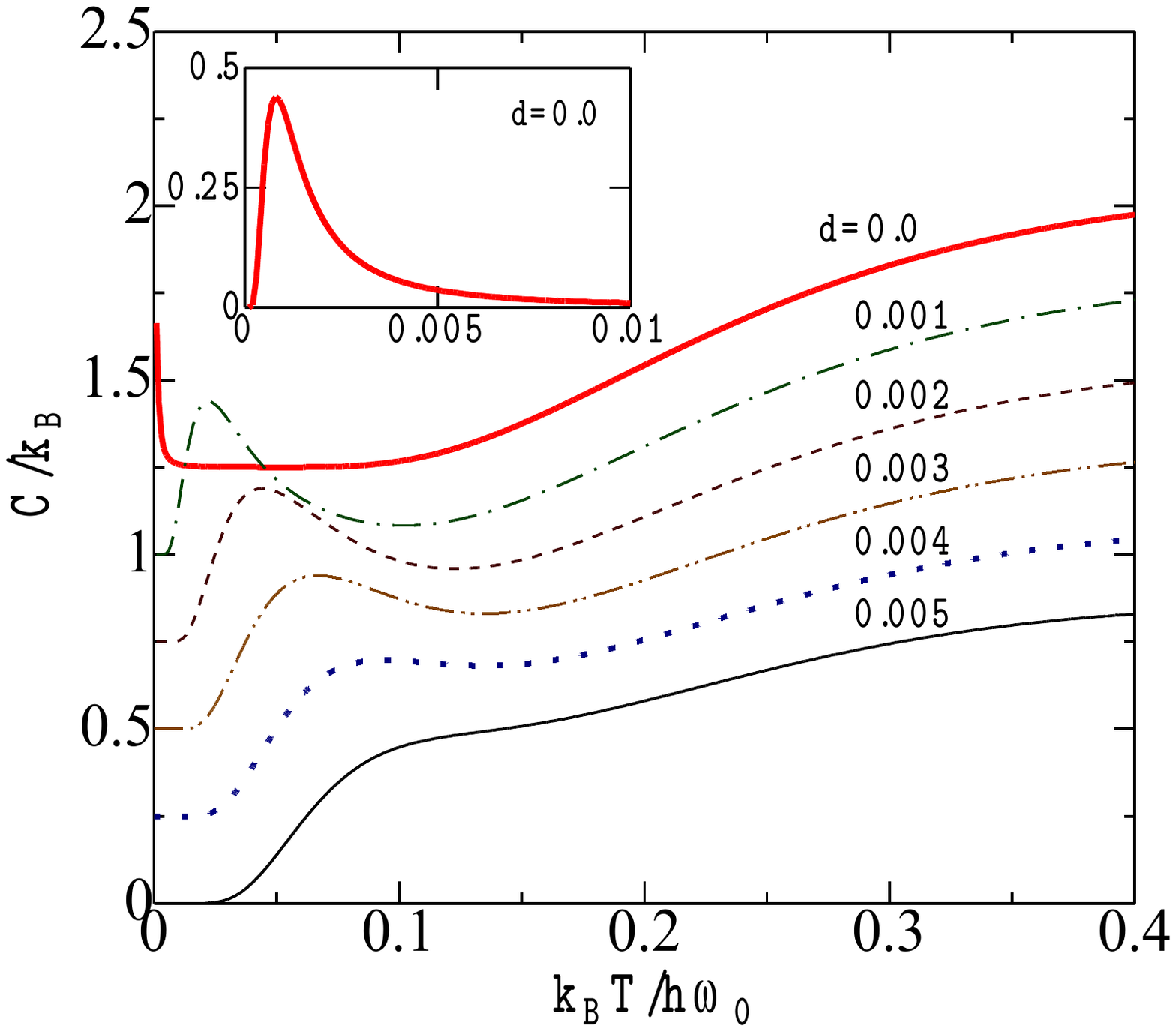}
\end{center}
\caption{
(Color online) 
The temperature dependence of the quantum specific heat $C(T)$ of model A with
the asymmetric potential [Eq. (\ref{eq:C1})] for various $d$: 
$d=0.0$ (bold solid curve), 0.001 (chain curve), 0.002 (dashed curve),
0.003 (double-chain curve), 0.004 (dotted curve) and
0.005 (solid curve), 
curves being successively shifted upward by 0.25 for clarity of figure.
The inset shows an enlarged plot of $C(T)$ with $d=0.0$ at $0 < k_B T/\hbar \omega_0 < 0.01$.
Note that the temperature dependence of $C(T)$ for a negative $d$ is the same as 
that for a positive $\vert d \vert$.
}
\label{fig6}
\end{figure}

Figure \ref{fig6} shows quantum specific heats calculated with asymmetric potentials
for various $d$ values \cite{Note3}.
The specific heat for $d=0.0$ has the Schottky-like anomaly at very low temperature
of $T \simeq 0.001$ (see the inset). 
When a small asymmetry of $d=0.001$ (or 0.002) is introduced, the position of 
the Schottky-type peak moves to higher temperature because of an increased gap of $\delta$.
For $d=0.005$, the Schottky-type peak almost disappears and its trace is realized as a shoulder
at $k_B T/\hbar \omega_0 \sim 0.1$. 
When we adopt a negative $d$, $\Delta U$ changes its sign but $\delta$ does not (Table 1). 
The temperature dependence of $C(T)$ for a negative $d$ with $\Delta U < 0$ 
is the same as that for a positive $\vert d \vert$ with $\Delta U > 0$.
Although the asymmetry has appreciable effects on the specific heat at low temperatures,
it has no effects at higher temperatures of $k_B T/\hbar \omega_0 \gtrsim 1.0$
for adopted asymmetry parameters. 


\subsection{A quadratic DW potential perturbed by Gaussian barrier (model B)}
\begin{figure}
\begin{center}
\includegraphics[keepaspectratio=true,width=80mm]{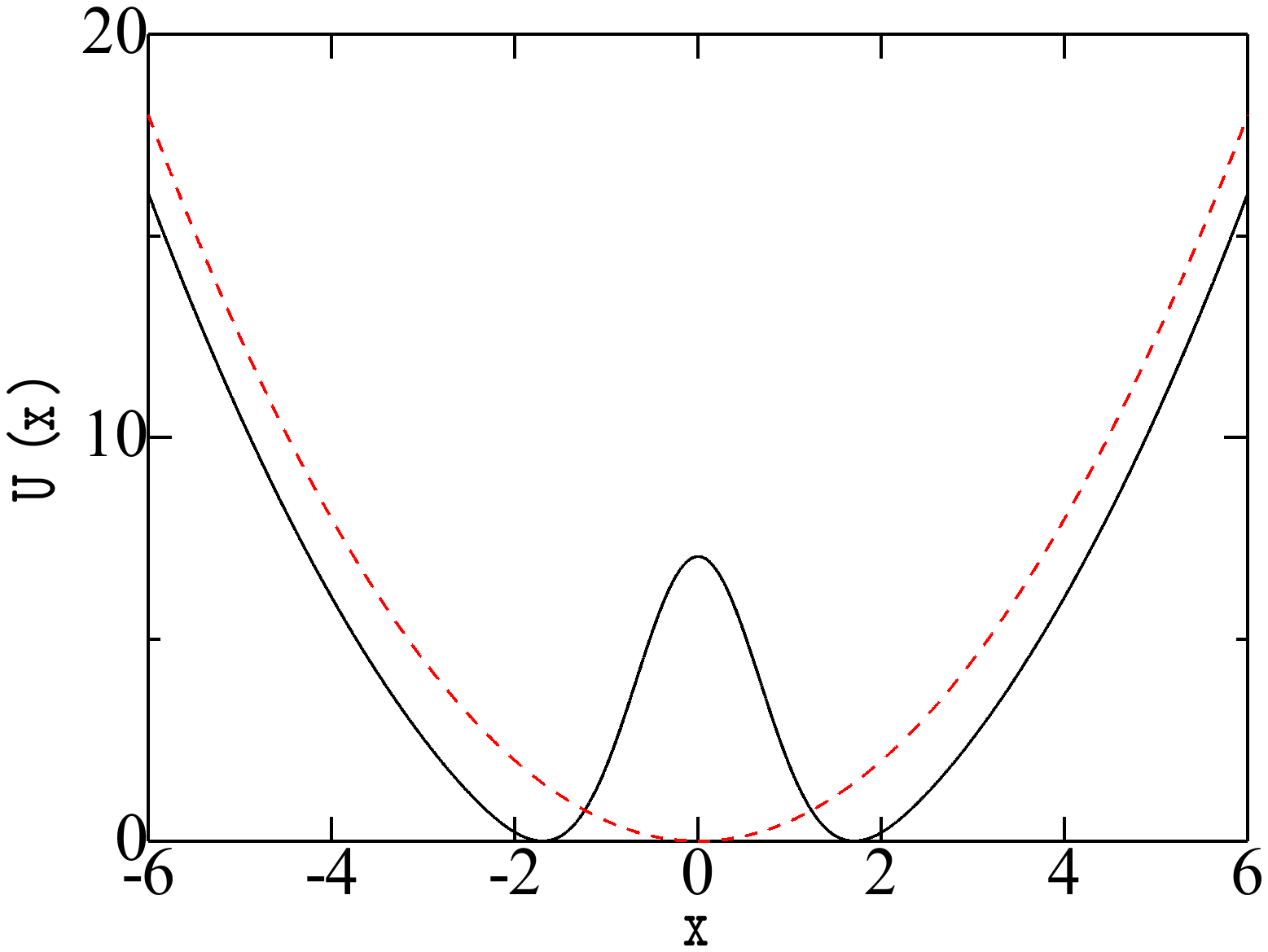}
\end{center}
\caption{
(Color online) 
The symmetric DW potential of model B [Eq. (\ref{eq:B1})] (solid curve) and  
the harmonic potential [Eq. (\ref{eq:B1b})] (dashed curve).
}
\label{fig7}
\end{figure}

We will apply our combined method to model B with a symmetric quadratic potential perturbed 
by a Gaussian barrier given by \cite{Chan63,Lin07}
\begin{eqnarray}
U(x) &=& U_0(x) +a \:e^{-b x^2}+c,
\label{eq:B1} \\
U_0(x) &=& \frac{m \omega_0^2 x^2}{2},
\label{eq:B1b}
\end{eqnarray}
where $a$ and $c$ are parameters.
The potential given by Eq. (\ref{eq:B1}) has  
stable minima at $x = \pm \sqrt{\ln (2a b/m \omega_0^2)/b} \equiv \pm x_0$
and an unstable maximum at $x=0.0$.
For our numerical calculations, we assume $m=1.0$, $\omega_0=1.0$, $a=9.0$, $b=1.0$, $c=-1.945$
and $x_0=1.700$, which yield $U(x_0)=0.0$, $U(0)=7.055$, $U^{''}(x_0)=5.78$, 
and $\Delta=U(0)-U(x_0)= 7.055$. 
The adopted potential is plotted by solid curve in Fig. \ref{fig7}
where dashed curve expresses the harmonic potential given by Eq. (\ref{eq:B1b}).

\begin{figure}
\begin{center}
\includegraphics[keepaspectratio=true,width=100mm]{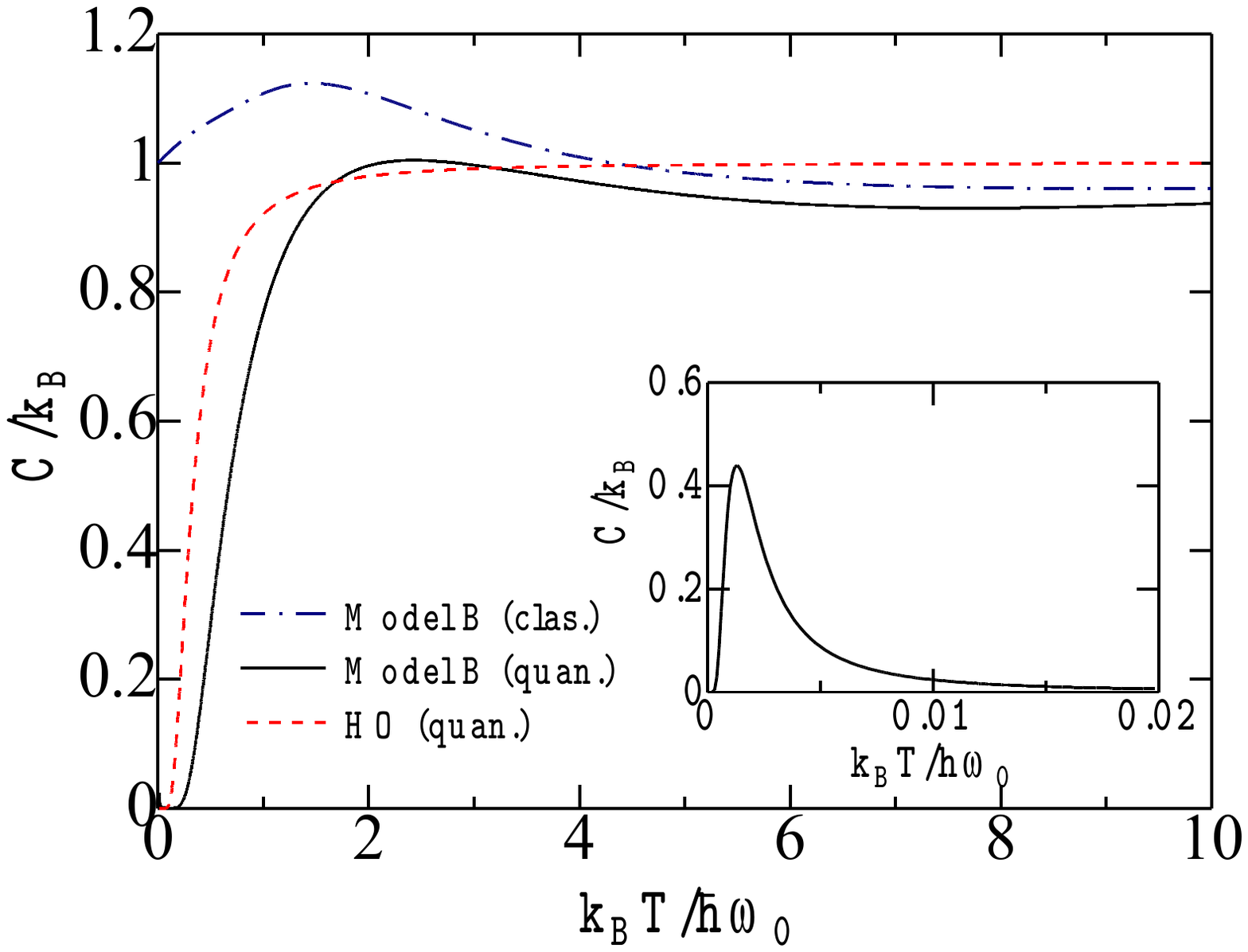}
\end{center}
\caption{
(Color online) Temperature dependences of
classical (chain curve) and quantum specific heats (solid curve) of model B
with the symmetric DW potential [Eq. (\ref{eq:B1})],
dashed curve expressing quantum specific heat of a harmonic oscillator (HO).
The inset shows an enlarged plot of the quantum specific heat at very low temperatures
with the Schottky-type anomaly.
}
\label{fig8}
\end{figure}

\begin{figure}
\begin{center}
\includegraphics[keepaspectratio=true,width=90mm]{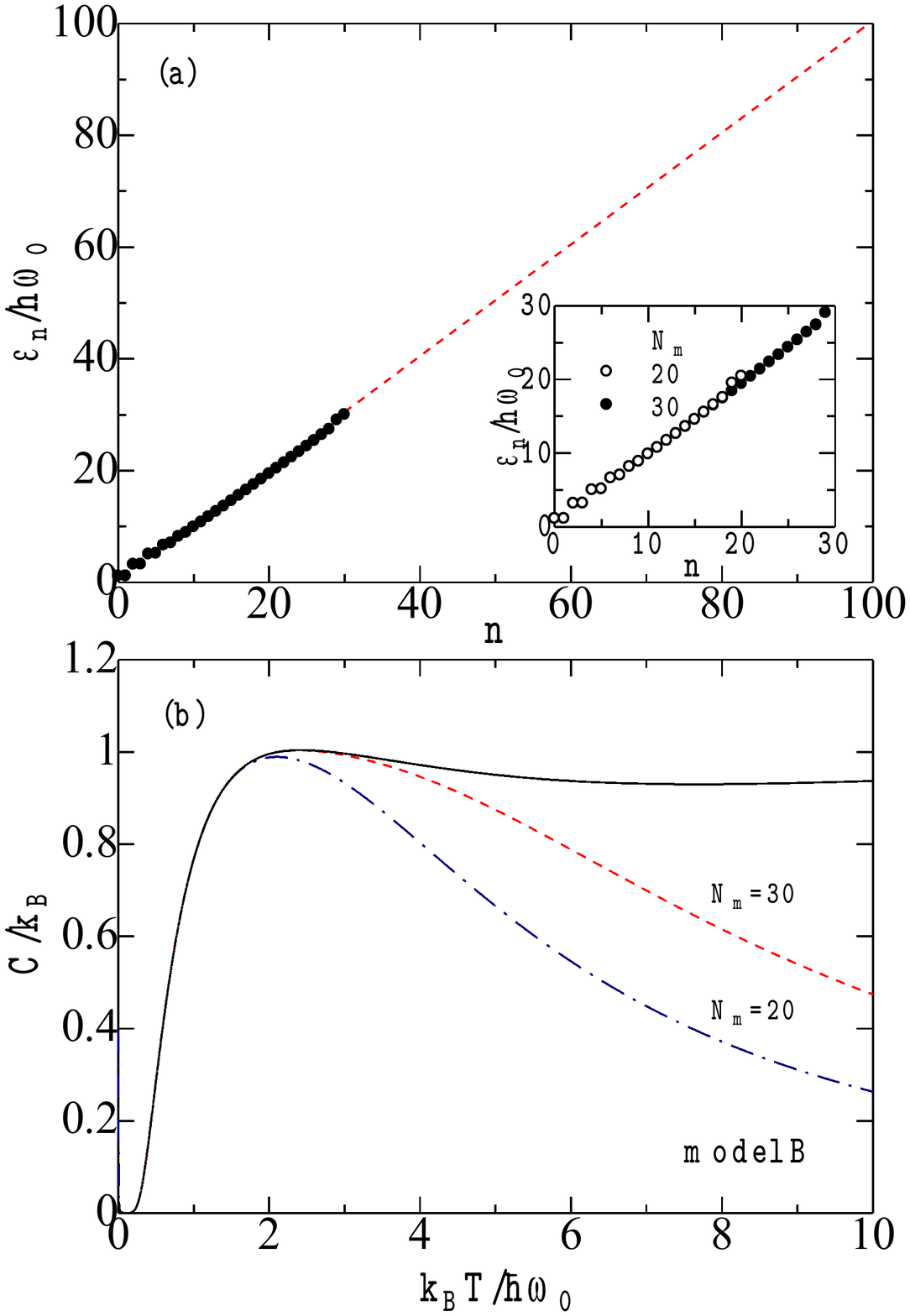}
\end{center}
\caption{
(Color online) 
(a) The $n$ dependence of eigenvalues $\epsilon_n$ of model B 
obtained by the energy-matrix diagonalization for $0 \leq n \leq N_m$ (circles) 
and by an extrapolation given by
$\epsilon'_n= (n+1/2) \hbar \omega_0$ for $n > N_m$ (dashed curve),
the inset showing eigenvalues obtained by the energy-matrix diagonalization with $N_m=20$ 
(open circles) and $N_m=30$ (filled circles).
(b) The temperature dependence of the quantum specific heat calculated 
with eigenvalues $\epsilon_n$ for $0 \leq n \leq N_m$ with $N_m=20$ (chain curve)
and $N_m=30$ (dashed curve), the solid curve expressing the result with
combined eigenvalues shown by the dashed curve in (a).
}
\label{fig9}
\end{figure}

\begin{figure}
\begin{center}
\includegraphics[keepaspectratio=true,width=100mm]{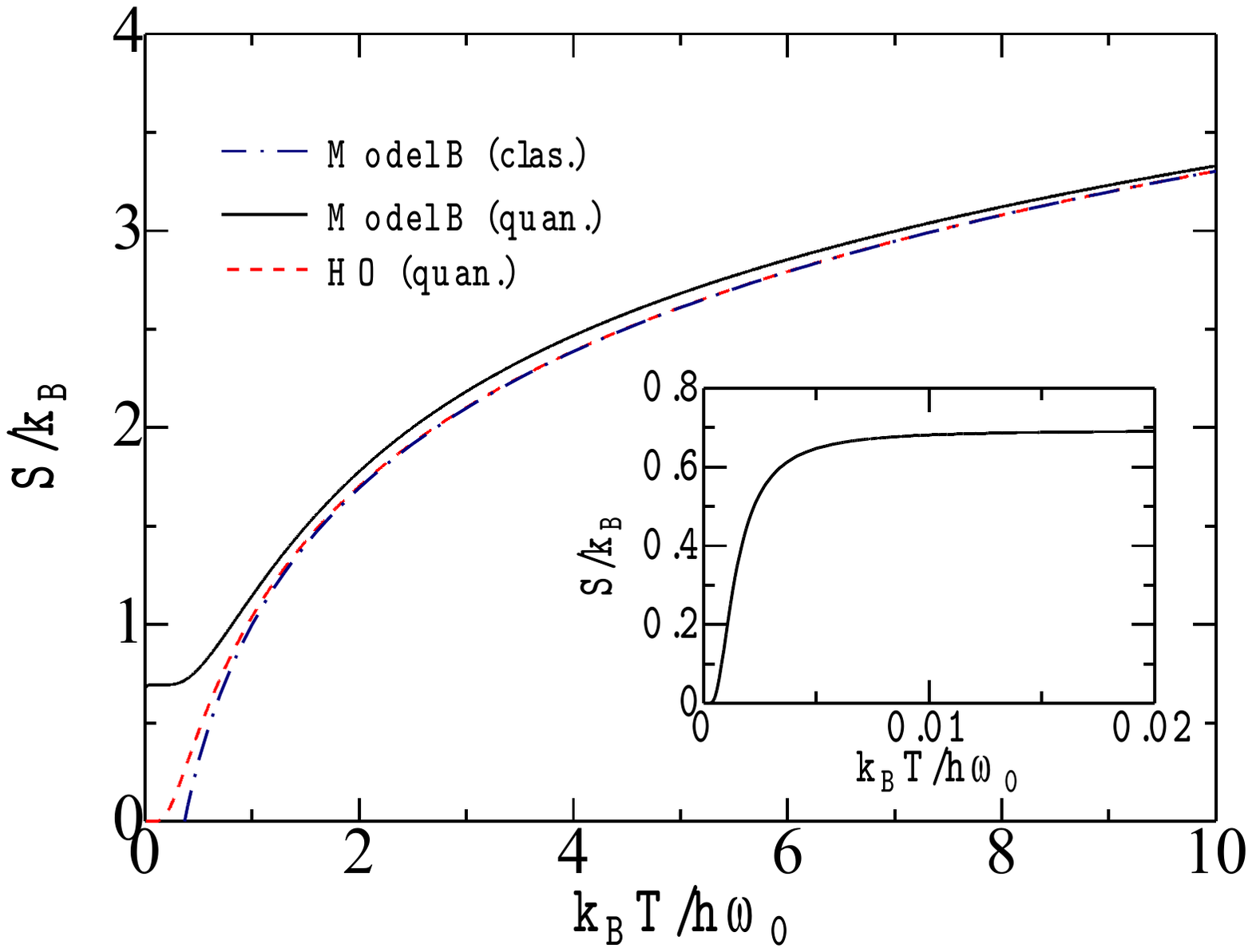}
\end{center}
\caption{
(Color online) Temperature dependences of
classical (chain curve) and quantum entropies (solid curve) of model B, and
the quantum entropy of a harmonic oscillator (HO),
the inset showing an enlarged plot of the quantum entropy at very low temperatures.
}
\label{fig10}
\end{figure}


We have numerically calculated the classical partition function 
to obtain the classical specific heat and entropy.
The calculated classical specific heat plotted by chain curve in Fig. \ref{fig8}
is not in good agreement with $C_{HO}$ of a harmonic oscillator 
at $k_B T/\hbar \omega_0 \lesssim 10$ although both reduce to $C = k_B$
in the high-temperature limit.
The calculated classical entropy will be explained shortly (Fig. \ref{fig10}).

For quantum statistical calculation, we have numerically evaluated eigenvalues
by the energy-matrix diagonalization. 
Matrix elements $H_{mn}$ of Eq. (\ref{eq:A14b}) are finite for any pair of even $\vert m-n \vert$.
Then the energy-matrix diagonalization for model B is more time consuming
than that for model A.
Eigenvalues $\epsilon_{n}$ calculated for $N_m=20$ and 30 are plotted 
in the inset of Fig. \ref{fig9}(a).
Eigenvalues for $n < 20 $ are almost the same for $N_m=20$ and 30.
Eigenvalues $\epsilon_{n}$ for $n=0$, 1, 2 and 3 are  
1.13021, 1.13332, 3.1931 and 3.21918, respectively, with $N_m=30$. 
$\epsilon_0$ and $\epsilon_1$ are quasi-degenerated with a small gap of
$\delta \equiv \epsilon_1-\epsilon_0=0.00311$.
We note that eigenvalues $\epsilon_n$ of model B in the inset of Fig. \ref{fig9}(a)
are similar to but slightly different from those of model A in the inset of Fig. \ref{fig3}(a).

Quantum specific heats calculated with the use of eigenvalues $\epsilon_n$
for $0 \leq n \leq N_m$ with $N_m=20$ and 30
are plotted by dashed and chain curves, respectively, in Fig. \ref{fig9}(b).
Both results with $N_m=20$ and 30 are in good agreement each other 
at $k_B T/\hbar \omega_0 \lesssim 2$ but significantly different
at $k_B T/\hbar \omega_0 \gtrsim 4$. 
We assume that extrapolated eigenvalues are given by
\begin{eqnarray}
\epsilon'_n &=& \left (n+\frac{1}{2} \right) \hbar \omega_0
\hspace{1cm}\mbox{for $N_m+1 \leq n < \infty$}.
\end{eqnarray}
The combined eigenvalues are shown by the dashed curve in Fig. \ref{fig9}(a).
The quantum specific heat calculated with the combined eigenvalues 
is shown by the solid curve in Fig. \ref{fig8} [or Fig. \ref{fig9}(b)].
The quantum specific heat has the Schottky-type peak at very low temperature at 
$k_B T/\hbar \omega_0 \simeq 0.0015 \sim \delta/2$, as shown in the inset of Fig. \ref{fig8}.
An increase of the quantum specific heat with raising the temperature
from zero is slower than that of
the harmonic oscillator plotted by the dashed curve in Fig. \ref{fig8}.
This is due to the fact that the curvature of $U(x)$ at the locally stable point
is larger than that of $U_0(x)$: $U^{''}(x_0) \; (=5.78) > U_0^{''}(0) \; (=1.0)$, 
which is realized in Fig. \ref{fig7}.
A comparison between Fig. \ref{fig8} and Fig. \ref{fig2} shows that although $C(T)$ of model B
is similar to that of model A at very low temperatures ($k_B T/\hbar \omega_0 \lesssim 0.02$), 
they are rather different at higher temperatures ($k_B T/\hbar \omega_0 \gtrsim 0.5$).
In the limit of $T \rightarrow \infty$, 
we obtain $C(T)=k_B$ in model B while $C(T)=(3/4)k_B$ in model A.

The temperature dependence of the classical and quantum entropies of model B
are shown by chain and solid curves, respectively, in Fig. \ref{fig10}, where
the quantum entropy of a harmonic oscillator is plotted by the dashed curve
for a comparison. 
The inset of Fig. \ref{fig10} shows the quantum entropy at very low temperatures.
With raising the temperature from zero, the entropy is rapidly developed to
$0.69$ ($\simeq \ln2$) at $k_B T/\hbar \omega_0 \simeq 0.005$, which is related 
with the Schottky-type specific heat 
at very low temperatures shown in the inset of Fig. \ref{fig8}.

\subsection{An asymmetric DW potential (FUK model)}
The specific heat of a DW system was calculated by FUK \cite{Feranchuk91} 
with the use of ZOM \cite{Feranchuk82} which is explained in the Appendix.
FUK adopted an asymmetric DW potential given by
\begin{eqnarray}
U_{FUK}(x) &=& \frac{1}{2} x^2 -\lambda x^3+ \gamma x^4,
\label{eq:C3}
\end{eqnarray}
with 
\begin{eqnarray}
\vert \lambda \vert > \lambda_m = \frac{4}{3}\:\sqrt{\gamma}, 
\label{eq:C3b}  
\end{eqnarray}
which has locally stable minima at $x=0$ and 
$x_s=(3 \lambda/8 \gamma) [1+\sqrt{1-16 \gamma/9 \lambda^2}]$,
and an unstable maximum at $x_u=(3 \lambda/8 \gamma) [1-\sqrt{1-16 \gamma/9 \lambda^2}]$.
Note that a prefactor of $x^2$ in $U_{FUK}(x)$ is positive which is required
for an application of ZOM \cite{Feranchuk82}, 
while that of the quartic DW potential given by Eq. (\ref{eq:C1}) is negative. 
The DW potential given by Eqs. (\ref{eq:C3}) and (\ref{eq:C3b}) becomes symmetric 
with respect to $x_u$ with $U_{FUK}(x_s)=U_{FUK}(0)=0$ and $U_{FUK}(x_u) =1/64 \gamma$ for
\begin{eqnarray}
\lambda= \lambda_c= \sqrt{2 \gamma}, \;\;\;
x_u &=& \frac{1}{2 \lambda},\;\;\;
x_s= \frac{1}{\lambda}.
\label{eq:D6}
\end{eqnarray}
For this symmetric case, a change of a variable $x$ with $u=1/2 \lambda$ leads to
\begin{eqnarray}
U_{FUK}(x+u)=\frac{(8 \gamma x^2-1)^2}{64 \gamma}, 
\end{eqnarray}
which is equivalent to $U(x)$ of model A in Eq. (\ref{eq:A4}).

\begin{figure}
\begin{center}
\includegraphics[keepaspectratio=true,width=80mm]{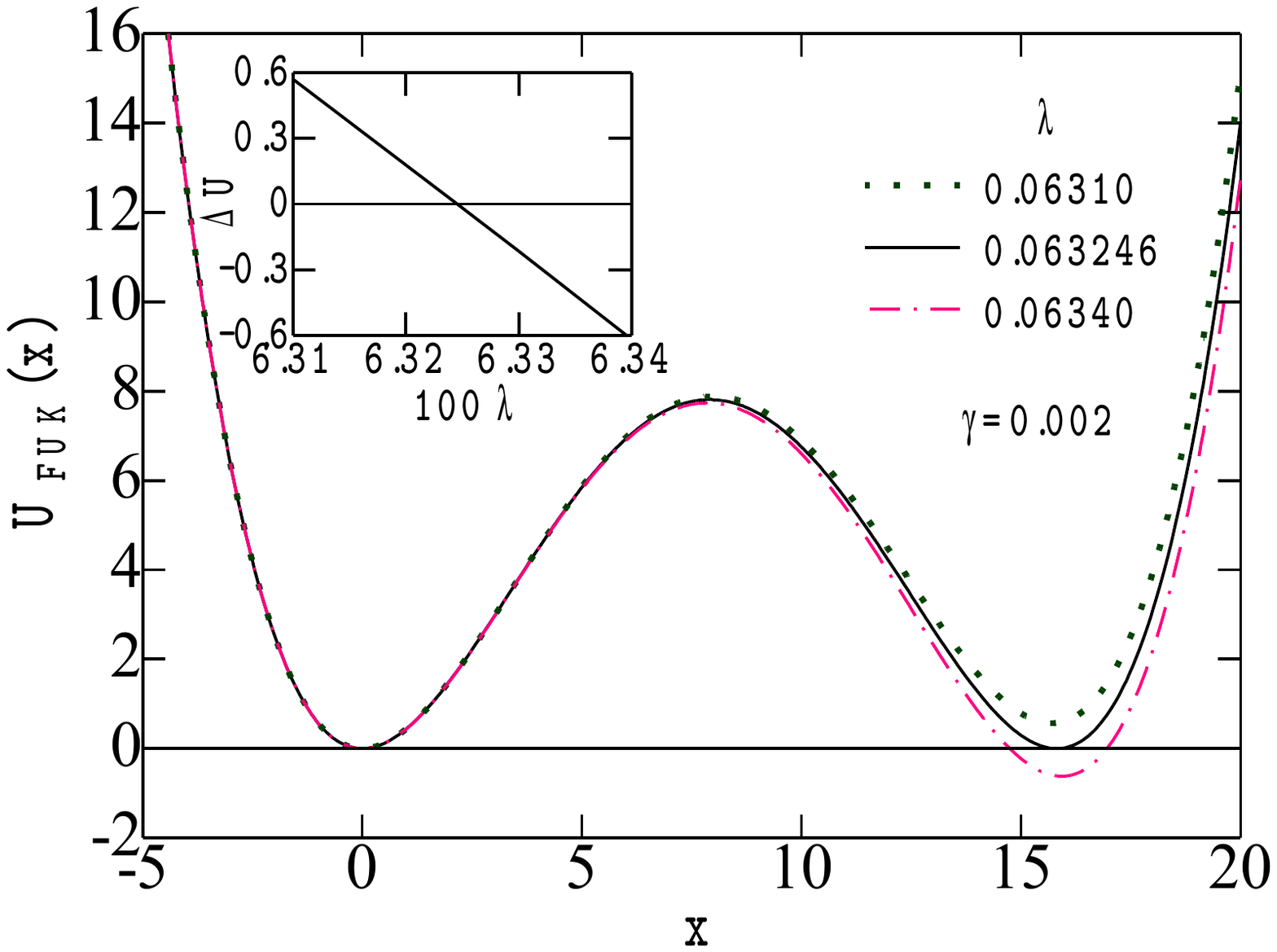}
\end{center}
\caption{
(Color online) 
The DW potential $U_{FUK}(x)$ given by Eq. (\ref{eq:C3}) for
$\lambda=0.06310$ (dashed curve), 0.063246 (solid curve)
and 0.06340 (chain curve) with $\gamma=0.002$, 
the inset showing $\Delta U$ [$=U_{FUK}(x_s)-U_{FUK}(0)$] 
as a function of $\lambda$ ($\times 100$).
}
\label{fig11}
\end{figure}

\begin{center}
\begin{tabular}[b]{|c|c|c|c|}
\hline
$\lambda$ &  $U_{FUK}(x_u)$  & $U_{FUK}(x_s)=\Delta U$ & $\delta$ 
\\ \hline \hline
0.06340 & $\;7.7370\;$ & $\;-0.61721 \;$ & 0.59848   \\
0.06335 &  7.7613  & $\;-0.41593 \;$ & 0.41286   \\
0.06330 &  7.7857  & $\;-0.21605 \;$ & 0.23789   \\
$\;0.063246\:(=\lambda_c)\;$ &  7.8125 & 0.0 & 0.11705  \\
0.06320 &  7.8351 &  0.17948 &  0.20636  \\
0.06315 &  7.8600 &  0.37514  & $\;0.37524\;$ \\
0.06310 &  7.8852 &  0.56939  & $\;0.55407\;$ \\
\hline
\end{tabular}
\end{center}

{\it Table 2} Potentials values at an unstable position ($x_u$) and a stable position ($x_s$), 
the potential difference $\Delta U$ [$=U_{FUK}(x_s)-U_{FUK}(0)$]
and the energy gap $\delta$ ($=\epsilon_1-\epsilon_0$) as a function of $\lambda$ 
with $\gamma=0.002$ for $U_{FUK}(x)$ [Eq. (\ref{eq:C3})] ($N_m=30$). 

\begin{figure}
\begin{center}
\includegraphics[keepaspectratio=true,width=100mm]{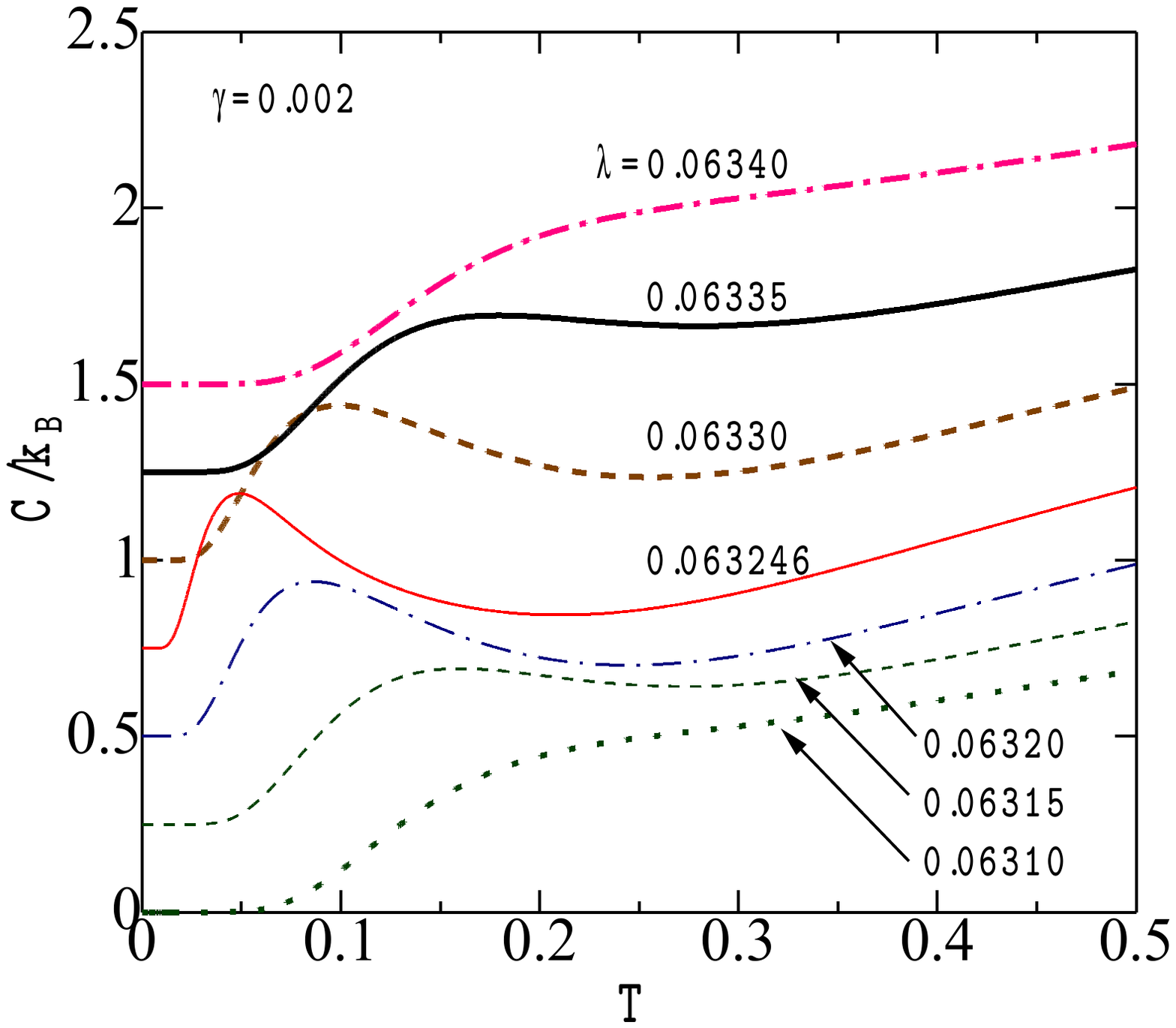}
\end{center}
\caption{
(Color online) 
Temperature dependences of the quantum specific heat of a DW system with 
$U_{FUK}(x)$ [Eq. (\ref{eq:C3})] for $\gamma=0.002$ with various $\lambda$:
$\lambda=0.06310$ (chain curve), 0.06315 (dashed curve), 
0.06320 (double chain curve), 0.063246 (solid curve), 0.06330 (bold double-chain curve), 
0.06335 (bold dashed curve) and 0.06340 (bold chain curve), curves being successively
shifted upward by 0.25 for clarity of figure.
}
\label{fig12}
\end{figure}

We have applied our combined method to a DW system with $U_{FUK}(x)$ 
given by Eq. (\ref{eq:C3}) with necessary modifications. 
We have chosen potential parameters of $\gamma=0.002$ and various $\lambda$ after FUK (see below).
The solid curve in Fig. \ref{fig11} expresses $U_{FUK}(x)$ for $\gamma=0.002$ and 
$\lambda=\sqrt{2 \gamma}\;(= \lambda_c \simeq 0.063246)$, which is
symmetric with respect to $x=7.9057$ ($=x_u=x_s/2$) with 
$U_{FUK}(x_s)=U_{FUK}(0)= 0.0$ and $U_{FUK}(x_u) =7.8125$.
When $\lambda$ is varied for a fixed value of $\gamma=0.002$, 
the potential difference between the two minima, $\Delta U \equiv U_{FUK}(x_s)-U_{FUK}(0)$,
changes: $\Delta U > 0$ ($\Delta U < 0$) for $\lambda < \lambda_c$ ($\lambda > \lambda_c$),
as shown in Table 2.
$U_{FUK}(x)$ for $\lambda=0.06310$ and $0.06340$ are plotted by
dashed and chain curves, respectively, in Fig. \ref{fig11} whose
inset expresses $\Delta U$ as a function of $\lambda$. 

By using the energy-matrix diagonalization, we have evaluated eigenvalues of $\{\epsilon_n \}$
for various $\lambda$ with $\gamma=0.002$ for $0 \leq n < N_m$ ($=30$). 
Table 2 shows that the calculated energy gap of $\delta$ ($=\epsilon_1-\epsilon_0$) is
minimum for $\lambda=\lambda_c$ and it is increased with increasing $\vert \lambda-\lambda_c \vert$.
By using obtained eigenvalues, we have calculated specific heats 
for various $\lambda$ whose results are plotted in Fig. \ref{fig12}.  
For $\lambda=\lambda_c \simeq 0.063246$ (solid curve) with $\Delta U=0$, the peak 
of the Schottky-type specific heat locates at $T \simeq 0.05 \sim \delta/2$. 
With decreasing $\lambda$ from $\lambda_c$, both $\Delta U$ and $\delta$ are increased, 
and then the peak position of the Schottky-type specific heat moves upward. 
For $\lambda \leq 0.06315$ the peak of the Schottky specific heat disappears, merging with the bump. 
On the other hand, with increasing $\lambda$ from $\lambda_c$, 
$\Delta U$ becomes negative as shown in Table 2. 
We note in Fig. \ref{fig12} that temperature dependences of $C(T)$ for
$\lambda=0.06330$ (bold double-chain curve), $\lambda=0.06335$ (bold dashed curve)
and $\lambda=0.06340$ (bold chain curve) are nearly the same as those
for $\lambda=0.06320$ (double chain curve), $\lambda=0.06315$ (dashed curve) and
$\lambda=0.06310$ (chain curve), respectively.  
Thus $C(T)$ of the FUK model is nearly symmetric with respect to $\lambda=\lambda_c$.
This is similar to the case of model A (Fig. \ref{fig6}) where $C(T)$ is symmetric
with respect to a degree of the asymmetry $d$.

\section{Discussion}
\subsection{A comparison between the results of FUK and ours}
The specific heat calculated for the FUK model shown in Fig. \ref{fig12} 
has been compared with $C_{FUK}(T)$ reported in Fig. 3 of Ref. \cite{Feranchuk91}.
A comparison between the two results shows that they are quite different in the following points:
(i) The Schottky-type anomaly of $C_{FUK}(T)$ locates at much higher temperature
with wider width than ours, 
(ii) $C_{FUK}(T)$ has more complicated temperature dependence than ours, and
(iii) the temperature dependence of our specific heat is almost symmetric
with respect to $\lambda=\lambda_c$, but $C_{FUK}(T)$ is not. As for the item (i), 
we suppose that it arises from a neglect of off-diagonal term $H_{od}$ in ZOM
[see Eq. (\ref{eq:D5}) in the Appendix]. 
Among matrix elements of the Hamiltonian, $H_{mn}$, given by
\begin{eqnarray}
H_{mn} &=& \langle m \vert H \vert n \rangle = \langle n \vert H_d \vert n \rangle \delta_{mn}
+ \langle m \vert H_{od} \vert n \rangle (1-\delta_{mn}),
\end{eqnarray}
only the first diagonal term is included in ZOM \cite{Feranchuk91} whereas 
both diagonal and off-diagonal terms for $0 \leq n, m \leq N_m$ are taken into account in our
numerical diagonalization method [Eqs. (\ref{eq:A14}) and (\ref{eq:A14b})]. 
We should note that the off-diagonal term plays an essential role in yielding a gap between
two stable states in the DW potential.
Indeed if off-diagonal contributions are neglected in our calculation, we cannot
obtain the Schottky-type specific heat. 

As for the item (ii), FUK claimed that the calculated, complicated $T$ dependence of $C_{FUK}(T)$ 
originates from the temperature-dependent energy gap \cite{Feranchuk91}.
This arises from the fact that optimum parameters of $\omega$, $u$ and $\nu$ 
given by Eq. (\ref{eq:D2}) in ZOM are determined at each temperature and 
then $\phi(\omega, u, \nu)$ in Eq. (\ref{eq:D2b}) is temperature dependent in general.  
It is natural that $C_{FUK}(T)$ is different from the Schottky-type specific heat 
which is obtained for the constant (temperature-independent) energy gap.

Related to the item (iii), FUK pointed out that $C_{FUK}(T)$ has a singularity when $\Delta U$ changes 
its sign \cite{Feranchuk91}. Such a result is, however, not realized in our calculation. 
As mentioned before, our $\lambda$-dependent $C(T)$ is almost symmetric with respect to $\lambda=\lambda_c$. 
On the contrary, the temperature dependence of $C_{FUK}(T)$ for $\lambda > \lambda_c$ is 
quite different from that for  $\lambda < \lambda_c$.

FUK \cite{Feranchuk91} reported the specific-heat calculation also for a different set of
parameters of $\lambda=0.01$ and $\gamma=0.002$ 
for which $U_{FUK}(x)$ has a single-minimum structure because
they do not satisfy the condition given by Eq. (\ref{eq:C3b})
($\lambda_m=0.0596$ for $\gamma=0.002$).
We have realized that $C_{FUK}(T)$ for these parameters (see Fig. 2 of Ref. \cite{Feranchuk91})
is in agreement with the specific heat obtained by our calculation (related results not shown).
It is suggested that although ZOM is not applicable to DW systems,
it may provide reasonable results for single-well systems where
off-diagonal contributions are expected to be unimportant. 
This is consistent with the fact that ZOM yields good results 
for systems with the anharmonic potential \cite{Feranchuk82} 
and the Morse potential \cite{Feranchuk88}. 

\subsection{Triple-well systems}

Finally we will study the triple-well system with the sextic potential
\begin{eqnarray}
U(x) &=& \frac{a x^6}{6}+ \frac{b x^4}{4}+ \frac{c x^2}{2},
\label{eq:C4}
\end{eqnarray}
with parameters $a$, $b$ and $c$, whose quasi-exact eigenvalues were investigated 
in Ref. \cite{Turbiner88}. 
When we concentrate our attention to low-lying three states with $U(-x_{s})=U(0)=U(x_{s})$, 
their eigenvalues are approximately given by
\begin{eqnarray}
\epsilon_0,\;\;\;\epsilon_1=\epsilon_0+\delta,\;\;\;
\epsilon_2=\epsilon_1+ \ \delta,
\end{eqnarray}
where $x=0$ and $\pm x_{s}$ express locally stable positions
and $\delta$ denotes energy gap between successive states.
At low temperatures, this energy spectrum yields the Schottky specific
heat whose peak locates at $k_B T=(3/4) \:\delta$.
On the other hand, at high temperatures, the sextic potential leads to
the classical specific heat given by $C=(1/2+1/6) k_B=(2/3) k_B$.
The sextic potential given by Eq. (\ref{eq:C4}) may be of single-, double- and triple-minima type,
depending on parameters of $a$, $b$ and $c$.
It would be worthwhile to investigate how thermodynamical properties of the sextic potential
system are changed against a change of the potential type, 
whose detailed study is left as our future subject.

\section{Concluding remark}
We have calculated specific heats of quantum DW systems with a quartic potential (model A), 
a quadratic potential perturbed by Gaussian barrier (model B) 
and $U_{FUK}(x)$ (FUK model) \cite{Feranchuk91}, by using the combined method 
in which eigenvalues obtained by finite-size energy-matrix diagonalization 
as well as extrapolated ones are included.
Specific heat and entropy in models A and B with symmetric potentials
have the Schottky-type anomaly at very low temperatures, which arises from low-lying
eigenstates with a small gap due to a tunneling through the potential barrier.
This is a quantum effect characteristic in DW systems, 
which is sensitive to an asymmetry in DW potentials.
In the high-temperature limit, specific heats of models A and B reduce to
$C=(3/4)k_B$ and $C=k_B$, respectively:
the former is different from $C_{HO}=k_B$ of a harmonic oscillator.

Advantages of our numerical combined method are that the calculations is physically 
transparent and that it yields correct results in both low- and high-temperature regions.
We have pointed out that the specific heat of DW systems calculated with
ZOM \cite{Feranchuk91} is incorrect because it neglects
off-diagonal contributions which play essential roles for a tunneling in DW potentials.
Although analytical methods such as the path-integral method (PIM) 
\cite{Feynman86,Kleinert06,Okopinska87} and the Gaussian wavepacket method (GWM) \cite{GWM} 
have been proposed to obtain the partition function of quantum DW systems, 
they are not suited for calculations of their specific heat \cite{Note2}. 
The present calculation has clarified
a basic problem on the specific heat of a DW system 
expressed by a pedagogical toy model, 
which is a basis for a study on more realistic DW systems, 
for example, described by system-plus-bath models \cite{Hasegawa11,Hasegawa12}.

\begin{acknowledgments}
This work is partly supported by
a Grant-in-Aid for Scientific Research from 
Ministry of Education, Culture, Sports, Science and Technology of Japan.  
\end{acknowledgments}

\appendix*

\section{Zeroth-order operator method}
\renewcommand{\theequation}{A\arabic{equation}}
\setcounter{equation}{0}
We will briefly mention ZOM \cite{Feranchuk88,Feranchuk91} in which
operators $p$ and $q$ are transformed by
\begin{eqnarray}
p &=& i \sqrt{\frac{\omega}{2}}(a^{\dagger}-a),\;\;\;
q =\frac{1}{\sqrt{2 \omega}}(a^{\dagger}+a)+u,\;\;\; 
[a, a^{\dagger}]=1,
\label{eq:D1}
\end{eqnarray}
with parameters $\omega$ and $u$,
$a^{\dagger}$ and $a$ denoting creation and annihilation operators, respectively.
Substituting Eq. (\ref{eq:D1}) to Hamiltonian given by Eq. (\ref{eq:C3}), 
we obtain
\begin{eqnarray}
H &=& H_d+H_{od},
\label{eq:D5}
\end{eqnarray}
where $H_d$ and $H_{od}$ are diagonal and off-diagonal parts, respectively,
with $[H_d, \hat{n}]=0$ and $\hat{n}=a^{\dagger} a$. Neglecting the off-diagonal term $H_{od}$,
we retain only the diagonal term $H_d$ in ZOM, 
\begin{eqnarray}
H_d \vert \psi_n \rangle &\simeq& E_n(\omega, u) \vert \psi_n \rangle,\;\;\;
\vert \psi_n \rangle \simeq \vert n \rangle,\;\;\;
a^{\dagger}a \vert n \rangle = n  \vert n \rangle,
\end{eqnarray}
where $\vert \psi_n \rangle$ and $E_n(\omega, u)$ denote approximate eigenfunction and eigenvalue, 
respectively, and $E_n(\omega, u)$ are expressed in terms of $\omega$, $u$ and $n$ 
(see Eqs. (6)-(8) in Ref. \cite{Feranchuk91}).
The partition function expressed by
\begin{eqnarray}
Z(\beta) &=& \sum_{n=0}^{\infty} e^{-\beta \epsilon_n}
\simeq \sum_{n=0}^{\infty} \langle \psi_n \vert e^{- \beta H} \vert \psi_n \rangle
\label{eq:D3}
\end{eqnarray} 
is transformed to \cite{Feranchuk82}
\begin{eqnarray}
Z(\beta) &=& \langle \nu \vert e^{-\hat{R} } \vert \nu \rangle,
\label{eq:D4}
\end{eqnarray} 
where operators $\hat{R}$, $\hat{k}$ and a state $\vert \nu \rangle$ are given by
\begin{eqnarray}
\hat{R} &=& \beta H+ \hat{k} \ln \nu + \ln N(\nu), 
\;\;\; N(\nu)=1- \nu, \\
\vert \nu \rangle &=& N(\nu)^{1/2} \sum_{n=0}^{\infty}
\nu^{n/2} \vert \psi_n \rangle,\;\;\;
\langle \nu \vert \nu \rangle = 1, \\
\hat{k} \vert \psi_k \rangle &=& k \vert \psi_k \rangle,
\end{eqnarray}
with a parameter $\nu$ ($\in [0, 1]$). 
By using the Bogoljubov inequality, $Z(\beta)$ is approximately calculated by
\begin{eqnarray}
Z(\beta)  & \geq & e^{-\beta \langle \nu \vert \hat{R} \vert \nu \rangle} 
= e^{-\beta \phi(\omega, u,\nu)} \equiv Z_0(\beta), 
\end{eqnarray}
where
\begin{eqnarray}
- \beta \phi(\omega, u, \nu) &=& -\beta N(\nu) \sum_{n=0}^{\infty} \nu^n E_n(\omega, u)
-\frac{\nu}{(1-\nu)} \ln \nu -\ln(1-\nu).
\label{eq:D2b}
\end{eqnarray}
Variational conditions given by
\begin{eqnarray}
\frac{\partial \phi(\omega, u, \nu)}{\partial \omega} 
&=& \frac{\partial \phi(\omega, u, \nu)}{\partial u}
=  \frac{\partial \phi(\omega, u, \nu)}{\partial \nu}=0,
\label{eq:D2}
\end{eqnarray}
yield self-consistent equations for optimum values of $\omega$, $u$ and $\nu$,
from which the approximate, optimized partition function $Z_0(\beta)$ may be obtained.
By using ZOM, FUK calculated the specific heat of a DW system 
with $U_{FUK}(x)$ \cite{Feranchuk91}.



\begin{thebibliography}{99}

\bibitem{Thorwart01}M. Thorwart, M. Grifoni, and P. H\"{a}nggi,
Annals Phys. {\bf 293}, 14 (2001).

%
\bibitem{Bagchi03}B. Bagchi and A. Ganguly,
J. Phys. A {\bf 36}, L161 (2003).

\bibitem{Manning35}M. F. Manning,
J. Chem. Phys. {\bf 3}, 136 (1935).

%
\bibitem{Razavy80}M. Razavy,
Am. J. Phys. {\bf 48}, 285 (1980).

\bibitem{Turbiner88}A. V. Turbiner,
Commun. Math. Phys. {\bf 118}, 467 (1988).

\bibitem{Caswell79}W. E. Caswell,
Ann. of Phys. {\bf 123}, 153 (1979).

\bibitem{Balsa83}R. Balsa, M. Plo, J. G. Esteve and A. F. Pacheco, 
Phys. Rev. D {\bf 28}, 1945 (1983).

\bibitem{Quick85}R. M. Quick and H. G. Miller,
Phys. Rev. D {\bf 31}, 2682 (1985). 

\bibitem{Turbiner09}A. V. Turbiner,
Int. J. Mod. Phys. A {\bf 25}, 647 (2010).

\bibitem{Chan63}S. I. Chan and D. Stelman,
J. Chem. Phys. {\bf 39}, 545 (1963).

\bibitem{Lin07}Chih-Kai Lin, Huan-Cheng Chang, and S. H. Lin,
J. Phys. Chem A {\bf 111}, 9347 (2007).

\bibitem{Feynman86}R. P. Feynman and H. Kleinert,
Phys. Rev. A {\bf 34}, 5080 (1986).

\bibitem{Kleinert06}H. Kleinert,
{\it Path Integrals In Quantum Mechanics, Statistics, Polymer Physics, And Financial Markets}
(Fourth Ed.) (World Scientific Pub. Co., Singapore 2006).
In Sec. 5 of this book, the variational perturbation theory of the PIM
is applied to DW systems.

\bibitem{Okopinska87}A. Okopi\'{n}ska,
Phys. Rev. D {\bf 36}, 2415 (1987).

\bibitem{Feranchuk91}I.D. Feranchuk, A.P. Ulyanenkov, and V.S. Kuz'min,
Chem. Phys. {\bf 157}, 61 (1991).

\bibitem{Feranchuk82}I.D. Feranchuk and L. I. Komarov,
Phys. Lett. A {\bf 88}, 211 (1982).


\bibitem{MATH}MATHEMATICA programs are available at URL:
http://www.lehman.cuny.edu/faculty/ \\
dgaranin/Mathematical\_physics-14-Eigenvalue\%20problems.pdf.

\bibitem{Note3}Numerical calculations of the quantum specific heat have been made
with the use of the relation: 
$C=k_B \beta^2 (\langle \epsilon_n^2 \rangle-\langle \epsilon_n \rangle^2)$
where $\langle \epsilon_n^k \rangle= \sum_n \:\epsilon_n^k \:e^{-\beta \epsilon_n}/Z$
with $Z= \sum_n e^{-\beta \epsilon_n}$, whose results are cross-checked with those
obtained by $C=k_B \beta^2 \partial^2 \ln Z/\partial \beta^2$.

\bibitem{Feranchuk88}I.D. Feranchuk and V. N. Tok,
Chem. Phys. Lett. {\bf 150}, 78 (1988).

\bibitem{GWM}
E. J. Heller, J. Chem. Phys. {\bf 62}, 1544 (1975).

\bibitem{Note2}In analytical methods such as OM, PIM and GWM,
the partition function $Z(\beta, \{ \mu \})$ is expressed in terms of a set of
variational parameters $\{ \mu \}$ whose optimum values are determined at
each temperature.  Numerical calculations of the specific heat which is expressed by
$C(T)=k_B \beta^2 \:\partial \ln Z(\beta, \{ \mu^* \})/\partial \beta^2$, are very difficult 
because optimized parameters $\{ \mu^* \}$ have the temperature dependence.

\bibitem{Hasegawa11}H. Hasegawa,
J. Math. Phys. {\bf 52}, 123301 (2011).

\bibitem{Hasegawa12}H. Hasegawa,
arXiv:1208.0295.

\end{thebibliography}
\end{document}